\documentclass[11pt,twoside]{article}
\usepackage{asp2014}

\aspSuppressVolSlug
\resetcounters

\begin{document}

\title{Escaping from the herd of white elephants}

\author{Jan~Noordam$^1$, Yan~Grange$^2$, and Peter~Teuben$^3$}
\affil{$^1$Madroon Community Consultants, De Marke 35, 7933 RA, Pesse, The Netherlands; \email{noordam@astron.nl}}
\affil{$^2$ASTRON, Netherlands Institute for Radio Astronomy, Oude Hoogeveensedijk 4, 7991 PD, Dwingeloo, The Netherlands}
\affil{$^3$Astronomy Department, University of Maryland, College Park, Maryland, USA}

\paperauthor{Jan~Noordam}{noordam@astron.nl}{ORCID}{Madroon Community Consultants}{}{Pesse}{The Netherlands}{7933 RA}{The Netherlands}
\paperauthor{Yan~Grange}{grange@astron.nl}{0000-0001-5125-9539	}{ASTRON}{Astronomy Group}{Dwingeloo}{}{7991PD}{The Netherlands}
\paperauthor{Peter~Teuben}{teuben@astro.umd.edu}{0000-0003-1774-3436}{University of Maryland}{Astronomy Department}{College Park}{MD}{20742}{USA}




\begin{abstract}
About 60 ADASS participants discussed the evolving roles of users, developers, and managers of astronomical data reduction software. Special emphasis was placed on the role of the user in the era of Big Data. Is it really inevitable that the increasingly ignorant user will just have to be satisfied with the standard products from data processing centres? 
\end{abstract}

\section{Introduction}

The idea of this BoF started as compassion. Over the last 40 years,
the quality of radio-astronomical observations has improved by up to 4
orders of magnitude. But for historical reasons, the data-reduction
software for producing these impressive results is scattered over
several different packages, each with its own user interface and user
support, and its own issues with installation and maintenance. All
this makes the software increasingly difficult to use. What's worse,
while exciting new algorithms are toppling the old show-stoppers, they
are slow to reach the user.

The present move towards pipeline processing addresses some of these
problems. Well-designed pipelines are the backbone of LOFAR programmes
like the LOTSS survey, or the detection of the Epoch of Reonization.
However, therein looms a totalitarian hazard: pipelines tend to be
black boxes that will be even less responsive to user comments than
the pioneering packages, which have profited so much from their active
involvement. Without the caring feed-back of the only people that can
judge the end-result, pipelines too will turn into White Elephants,
and stop evolving.

During preparation, the subject of this BoF evolved from compassion
with the long-suffering users of our data reduction software, to the
role of the scientific user in the era of Big Data.  The new
generation of giant radio telescopes (LOFAR, SKA) will need even more
sophisticated processing to fully exploit their dazzling but expensive
capabilities. But the consensus seems to be that data volumes will be
too large to be moved, while self-propelled users will not have access to
the necessary computing resources. So the question is whether it is
really inevitable that (increasingly ignorant) users will just have to be satisfied with the
standard products from data processing centers, or whether ways can be
found to involve them somehow?

\section{About the Discussion}

About 60 people attended this BoF, despite the fact that it clashed
with drinkies at the end of a long day. Because we only had an hour to
deal with a rather large subject, we limited ourselves to three
"sociological" Discussion Points about the roles of users and
developers. We stayed away from technical issues.

These three points were discussed with gusto. However, since there tend to
be few hard-core users at ADASS, and developers are thinkers rather
than talkers, the debate was somewhat dominated by software managers
and other generalists. Perhaps next year we should add a fourth
Discussion Point about their role.

Unfortunately, our audio recording failed us, so we have endeavoured
to reconstruct the essence from our poor short-term memories.

\subsection{Discussion Point 1: A new eco-system for users?}

It is clear that it will be harder for users to be involved in data processing in the era of Big Data. But that should not mean that they should be cut out of the loop.

For instance, users could still develop their own pipelines on their laptops, and exercise them on a succession of increasingly large subsets of their huge data
volumes stored in a Processing Centre. The well-proved pipeline could
then be run on the large computers of the Centre.

In this way, scientific users can still be actively involved in the
development of data-processing software.

Several times in the discussion it became apparent that we are not using 
our resources wisely (or read: we are suffering from a lack of resources).
We are also not using our outreach (e.g. documentation) effectively.  


\subsection{Discussion Point 2: A new eco-system for developers?}

There is a perceived disconnect between algorithm development and implementation and 
the involvement of computer science professionals/students.

Software packages used to be under the sole control of a single
developer. Subsequent groups of more than one developer required the added complication of Management, which changed the dynamic rather fundamentally.  Should
we now develop a system in which the work of (many) external developers can be
included, and offered to users in an accessible way? The associated
reward system (money?) could have interesting effects on competition
and maintenance.

\subsection{Discussion Point 3: user-developer Interaction?}

Accidental "Golden Teams" of one opinionated developer and one active
scientific user have been very important in the process that has
improved the quality of radio-astronomical observations by four(!)
orders of magnitude over the last 40 years or so.

This concept resonated strongly in the BoF discussions. It was
suggested that, in addition to user-developer teams, there should also
be Golden Teams of Mathematicians and Implementers. The problem then
becomes one of creating the right conditions, in which compatible(!)
users and developers can find each other, and get the opportunity
(freedom) to do their thing.

\section{Conclusion}

A tentative conclusion could be that scientific users do have an
important role to play, even in a time of Big Data, and that it would
be a mistake to give up on them.

\bibliography{B3}

\appendix

\section{Technical Solutions}

This BoF was explicitly limited to "sociological" issues, like the
need for making our data reduction software more accessible, and the
desirability of user control, even in the era of pipelines and Big
Data. However, the discussion of such issues will be taken more
seriously if there is a perception that viable technical solutions
exist to implement such requirements. 

It may also help to know that such discussions are not new within
ADASS. During the 1990's, the last afternoon of the conference used to
be devoted to a free-ranging exchange on the Future of Astronomical
Data analysis Systems \citep[FADS;][]{fads2}. After a brief introduction, the floor
would be opened up, after which the only problem of the moderator was
to gently curtail some speakers in order to give others a chance.

One of the ideas raised at the FADS was a triangular "Gaming Table", a
system in which Software Agents would link a given data structure(1)
with selected data operations(2), as specified by means of a user
interface(3) of choice. The highly competent Agents would hide all the
detail that a user should never have to see, thereby addressing the
two sociological issues mentioned above.

A quarter century later, these issues are still with us, as the
(unfortunately unrecorded) discussions at this BoF have demonstrated
so clearly.  
Fortunately, there are many ways forward. At the simplest level, existing legacy software that is terribly useful can still have a golden 
future if interfaces to modern scripting languages are available. An example of this has been implemented in the AMUSE software \citep{2018Sci...361..979P}.

     Another more recent idea is the introduction of a layer of
     "Proxy" interface objects between users and software modules.
     Such Proxies would take care of input arguments and output
     results, and all the access incantations that a user should never
     have to see. Because of their uniform structure, the Proxies can
     talk to each other, and be handled by a single GUI. It turns out
     that such an extra layer has many advantages, not only in
     offering users access to software and involving more developers,
     but also in providing structure to distributed collaborations.

But the good news is that we now have the technology to
implement solutions like the Gaming Table. As Ronald Reagan said: "You ain't seen nothing yet".


\end{document}